# On the Empirical Association between Trade Network Complexity and Global Gross Domestic Product

Mayank Kejriwal[1] and Yuesheng Luo[1]

University of Southern California, Los Angeles CA 90292, USA,
{kejriwal,luoyuesh}@isi.edu,
WWW home page: https://usc-isi-i2.github.io/kejriwal/

**Abstract.** In recent decades, trade between nations has constituted an important component of global Gross Domestic Product (GDP), with official estimates showing that it likely accounted for a quarter of total global production. While evidence of association already exists in macro-economic data between trade volume and GDP growth, there is considerably less work on whether, at the level of individual granular sectors (such as *vehicles* or *minerals*), associations exist between the complexity of trading networks and global GDP. In this paper, we explore this question by using publicly available data from the Atlas of Economic Complexity project to rigorously construct global trade networks between nations across multiple sectors, and studying the correlation between network-theoretic measures computed on these networks (such as average clustering coefficient and density) and global GDP. We find that there is indeed significant association between trade networks' complexity and global GDP across almost every sector, and that network metrics also correlate with business cycle phenomena such as the Great Recession of 2007-2008. Our results show that trade volume alone cannot explain global GDP growth, and that network science may prove to be a valuable empirical avenue for studying complexity in macro-economic phenomena such as trade.

## 1   Background and Motivation

In recent years, globalization has been a much discussed topic in both economic and political circles [20], [12]. On the political front, globalization has sharply divided many observers, often along the lines of political affiliation, with some claiming that it has led to millions of people in vulnerable countries escaping poverty [15], while others have pointed out the damage to jobs and real wages (especially in sectors like manufacturing) in developed countries [5]. At the heart of this issue lies trade. Over the last decade, estimates show that trade accounted for approximately a quarter of total global production [19]. Trade transactions include both goods and services, the latter including tourism and financial services, which do account in trade statistics and represent a flow in foreign capital. However, trade in goods is predominant from a global perspective, and production chains for these goods have become complex over time [18]. This complexity can also lead to fragility and supply chain disruption, as the pandemic recently showed [1].

Understanding the dynamics of trade, and its effect on global Gross Domestic Product (GDP), is a complex issue that is of great interest in macro-economics [7], but which has not fully been addressed from a *network science* perspective. In part, the issue arises because constructing a rich network over time, and across multiple important sectors (such as agriculture and machinery), requires data that takes time to gather and process. Recently, such data was released as part of the ongoing research being done by the Atlas of Economic Complexity project at Harvard University [10]. Although the primary goal of this project is data visualization that allows research into the exploration of global trade flows across markets, the project has released fine-grained data that also allows us to construct sector-specific trade networks between countries over the last 25 years, leading up to a year before the pre-pandemic period. This period not only covers the financial crisis of 2007-2008 [9], but also China's entry into the World Trade Organization, and its emergence as a global economic force [21].

In this paper, we consider a simply stated, but empirically complex, research agenda: can network science, as applied to global trade networks between countries, be used to study the association between global GDP growth and increased trade complexity? Note that there is already evidence of such association in macro-economic data between trade *volume* and GDP growth [19]. However, complexity is not a phenomenon that can be studied through correlations between two simple time-series of trade volume and GDP. Instead, we need to first construct sector-specific trade networks over time (e.g., per annum) and empirically assess if changes in selected network science metrics, such as average degree and density, correlate with equivalent changes in global GDP growth. Theoretically, we note that changes in the global trade network (at least, in an unweighted representation) are not necessary for achieving higher volume or higher GDP. For example, one can imagine an extreme situation where the trade network that existed in 1995 continues to exactly be replicated by the trade network in 2015, and global GDP growth is explained purely through volume growth. In other words, if there is no *structural* change in the network at all, average degree and density (among a host of other network science metrics) would not change, and volume alone could explain trade's component in global GDP growth.

However, our results show that there has been considerable structural change in trade networks, with rapid changes occurring in the period of high growth leading up to the 2007-2008 financial crisis, and flattening soon after. There are interesting correlations between sectors: some sectors tend to be recession-proof while others are heavily influenced. In an absolute sense, the changes are small; however, viewed at the appropriate scale, are marked and significant.

In the economics literature, there has been considerable recent interest in applying network science to the study of complex phenomena [14], [13]. Many such studies have involved the stock market, including modeling the stock market as a 'temporal' network; see, for example, representative work in [24], [25]. Others have used network science to understand risk in the interbank system [22], among other complex systems research with applications in economics [8]. There has also been considerable literature on studies of economic complexity [11], [2], [17]. Although preliminary, the study presented in this paper adds to this knowledge by considering the correlation between trade complexity (as viewed and measured from a network science lens) and global GDP. To the best of our knowledge, such a study (even in a preliminary form) has been lacking in the literature thus far. Other complex phenomena have had a long history of being investigated through network science; numerous examples are presented in numerous books and reviews on the subject [23], [6], [4], [3].

The rest of this paper is structured as follows. First, we detail materials and methods, including the manner in which the sector-specific and per-year trade networks were constructed and interpreted. Next, we report the findings of the study. Finally, we conclude the paper with a brief summary and some guidance on future research avenues.

## 2   Materials and Methods

### 2.1   Data

The data were obtained from the *Atlas*[1] project by Harvard's Growth Lab. Atlas includes trade statistics for 250 countries and regions, divided into 20 goods- categories and 5 service-categories. In total, 6,000 products are covered globally across the 25 categories. The raw trade data is collected from United Nations Statistical Division (COMTRADE) for goods, and the International Monetary Fund (IMF) for services. In this paper, we focus only on goods, which is described by two trade classification systems - Harmonized System (HS) 1992 and Standard International Trade Classification (SITC) revision 2.

HS data provides detailed classification of goods and their trade volumes between countries between 1995 to 2018. It covers about 5,000 goods across 10 sectors (Table 1). HS categories are broken down into 1-, 2-, 4-, and 6-digit detail levels. Each level provides data for goods at a finer granularity. For example, the 1-digit level, which is considered for the studies in this paper, describes the 10 sectors in Table 1. SITC data provides a longer time-series from 1962 to 2018 but has far lower goods coverage (only about 15% of goods covered by HS), since many goods were not invented yet or had significant trade volumes before the 1990s (an example being the personal computer or laptop). SITC categories are also broken down into 1-, 2-, and 4- digit detail levels.

For each of the two datasets, key variables include the country, trading partner, product and year. There is also detailed information on many other aspects of the trading relation, including export and import values (for a given good being traded between the two trading partners). While some of this information is used in our experiments, the primary information that we rely on for network construction (as subsequently described) is the set of goods being traded across countries over the time period 1995-2018 that is covered by the HS data.

---

[1] https://atlas.cid.harvard.edu/data-downloads

**Table 1.** The ten sectors covered in the HS dataset at the 1-digit level. Networks are constructed at this level of granularity, with one network typically constructed per sector as discussed in Section 2.2. The 'Other' sector is excluded from study due to its exceptional status.

| Sector | # 4-Digit Products | Examples |
|---|---|---|
| Textiles | 181 | Cotton sewing thread, globes |
| Agriculture | 290 | Crustaceans, peanuts |
| Stone | 67 | Gold, glass fibers |
| Minerals | 67 | Quicklime, nickel ore |
| Metals | 157 | Copper mattes, handsaws |
| Chemicals | 219 | Phosphides, sulfides of nonmetals |
| Vehicles | 38 | Railway service vehicles, wheelchairs |
| Machinery | 174 | Vacuum cleaners, thermionic |
| Electronics | 48 | Liquid crystal devices, games |
| Other | 2 | Trade data discrepancies, commodities not specified according to kind |

## 2.2  Network Construction and Measures

As a first step toward network construction, we isolated the Country-Partner-Product-Year variables at the 4-digit product level (HS system). In total, the combined file has more than 113 million rows. We filtered the countries to only keep the ones that have data through all time periods, with 213 out of 250 locations being retained. For a specific product, or a union of products[2], we build an undirected graph for each year using the networkX[3] package, by first constructing a node for each country involved in the trading of the selected products (all 4-digit level products falling within a given 1-digit level sector), and linking two countries with an edge if a trading relation (i.e. import or export of the sector-specific goods being studied) exists between those countries. One graph, called a *trade network*, is constructed per sector. Six graph attributes are calculated and compared over time: number of nodes, average node degree, average node clustering coefficient, density, total number of triangles, and diameter.

---

[2] Since we construct graphs at the sector level (1-digit HS level, a union of products is involved since the data is available at the 4-digit product level, and multiple 4-digit level products comprise a 1-digit level HS 'sector' (similar to a taxonomy).

[3] https://networkx.org/

## 3   Results

Figure 1 illustrates the average degree over time of the trade networks constructed for each of the five sectors, namely *Textiles, Agriculture, Stone, Minerals,* and *Metals*. A similar trend is observed for the other sectors (not shown herein). We find that the average degree of the network rapidly increases (by almost 50%) over a 15 year period (1995-2010) for all sectors and that the trends are largely consistent. Following the 2008 financial crisis, there is a plateauing effect, but it is unclear if this is due to long-term shifts due to the crisis itself or because of diminishing returns.

Figures 2, 3 and 4 plot some other common network metrics over time, such as average clustering coefficient, average density and total number of triangles, respectively. For the latter two, we see a near identical trend; however, it is important to note that the y-axis of the plots do not start at 0 for any of the plots. In comparing the numbers in 1995 and 2010, we find that the total number of triangles approximately doubles during that time, while the density increases by the same extent as the average degree. The metrics all suggest that significant increase in global trade was witnessed during that time. Importantly, this cannot be due only to the influence of a single country (such as China) or even a small set of countries, since we are not accounting for volume of trade in these networks. Since nodes in these networks are countries, and edges are only created if a minimum level of trade took place in a given sector between a pair of countries (as discussed earlier in *Materials and Methods*), each trade relation has a minimal degree of strength for it to be included as an edge in the network.

In contrast, the average clustering coefficient in Figure 2 exhibits a more cyclical trend but only when viewed at a sufficiently fine-grained level (the y-axis has a range of only about 0.04 units at the maximum). Otherwise, the curves would look mostly flat, and tend to range between 75-85% depending on the sector. We do note that in the period leading up to 2010 (i.e. in the aftermath of the 2008 recession, which is when the ramifications on trade would tend to be felt more) there is usually a significant decline, especially in the electronics, textiles, metals and vehicles sectors. In contrast, sectors such as agriculture and machinery, which tend to be capital resources that are needed in the manufacturing of a wide range of products (or in the case of agriculture, for servicing a basic need, such as food), the decline is less significant and continues to show an upward trend soon after. For electronics, we also see a similar precipitous decline in the immediate aftermath of the dot-com bubble in 2000. Hence, to a certain degree, cyclical economic fluctuations can be witnessed in network metrics, albeit with some lag (in other words, it would not be prudent to use such measures as predictive tools, only as analytical and quantitative measures of how trade is systemically affected by business cycles).

Although the plots are useful visual tools, we seek to better understand the correlation between these measures and the global GDP in Table 2. The table shows that there is indeed a high correlation (well above 0.9) for the series representing average degree, average density and total number of triangles, and far less of a correlation for the average clustering coefficient. Correlations for electronics, vehicles, chemicals, textiles and agriculture are most positive, and explain the critical role that trade and globalization plays in ensuring that these products are available at low prices for industries and markets worldwide. The role of agriculture in trade has not always been appreciated in recent decades; however, the ongoing war in Ukraine is already leading to commentary that food and agriculture (especially wheat) will be heavily affected as a consequence [16].

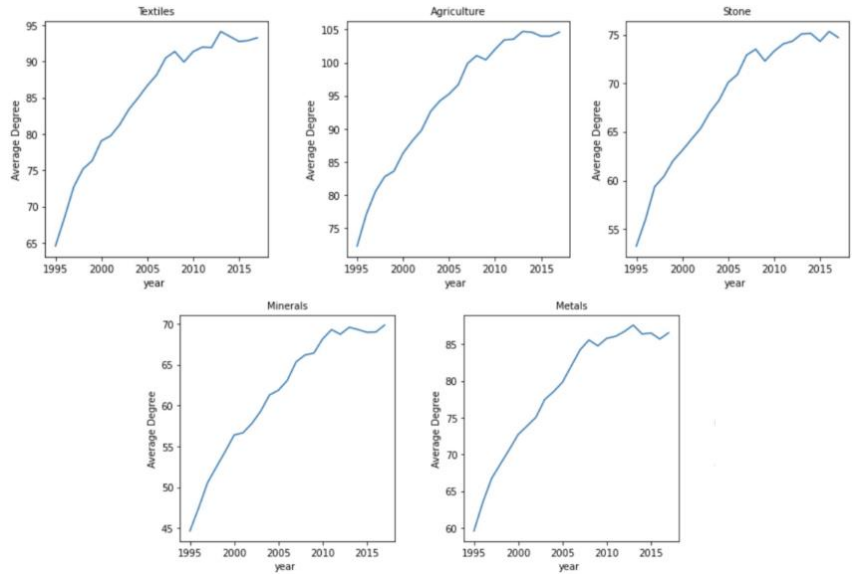

**Fig. 1.** Average degree over time of the trade networks constructed for the five sectors [Textiles, Agriculture, Stone, Minerals, Metals]. Distributions for the other four sectors were found to be qualitatively similar.

Finally, Figures 5 (a) and (b) show the average and percentage change, respectively, in trade volume (sum of exports and imports), with average taken over countries over the time period 1995-2018. For Figure 5 (a), change in trade volume for a certain country is calculated from the difference in sum of export and import between a current year and its previous year. For Figure 5 (b), percentage change is calculated by dividing change in trade volume with trade volume of the previous year[4].The (percentage) change in trade volume was then averaged over all countries. In analyzing the results, we find, concurrent with the previous figures, that trade volume indeed declines precipitously leading up to, and shortly in the aftermath of the financial crisis, but rebounds briefly (perhaps due to long-duration contracts that needed to be fulfilled), before continuing its decline all the way till 2015. There is some growth thereafter, most likely attributable to overall economic growth but a plateauing effect can already be observed shortly thereafter. The percentage change in Figure 5 (b) shows that trade has remained largely stable year-over-year but there has been marked fluctuation since the financial crisis of 2008[5]. In the wake of populist movements since 2015, the prospects of global trade witnessing a revival seem uncertain at best.

---

[4] Theoretically, it is possible for the trade volume of the previous year to be 0 but this never occurred in practice in the dataset.

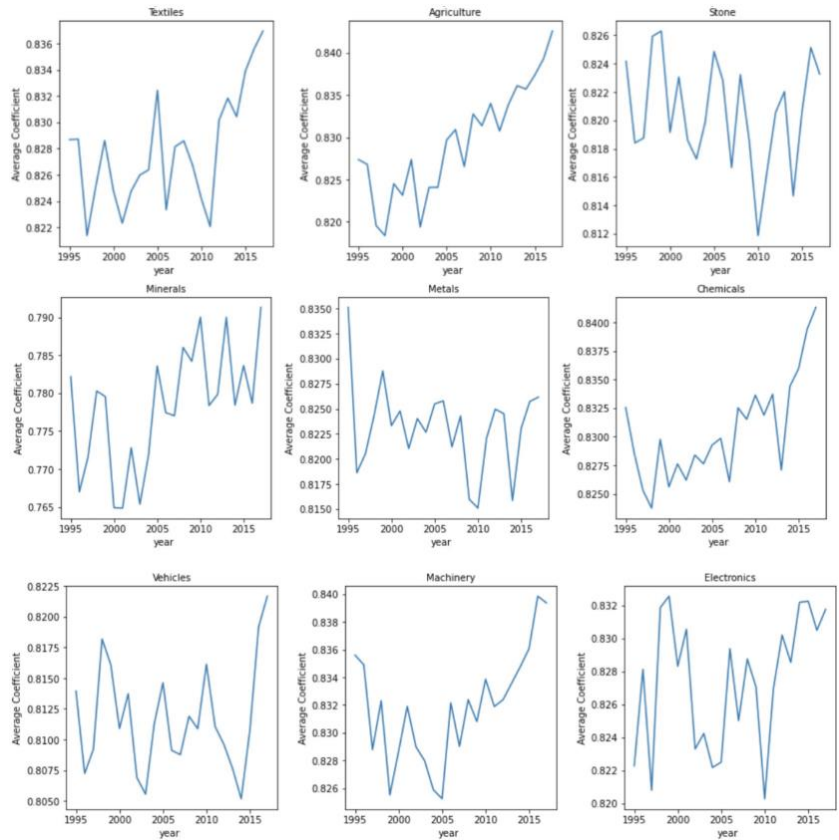

**Fig. 2.** Average clustering coefficients of nodes in the trade networks constructed for all nine sectors.

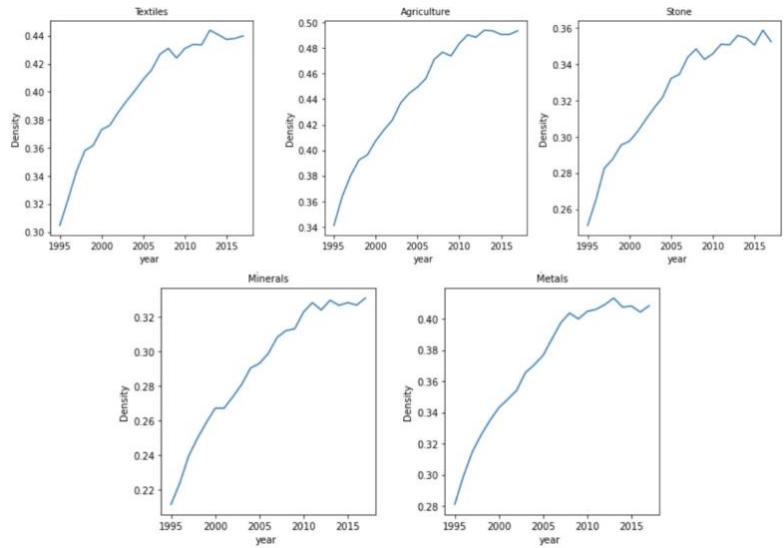

**Fig. 3.** Average density of the trade networks constructed for the five sectors [Textiles, Agriculture, Stone, Minerals, Metals]. Distributions for the other four sectors were found to be qualitatively similar.

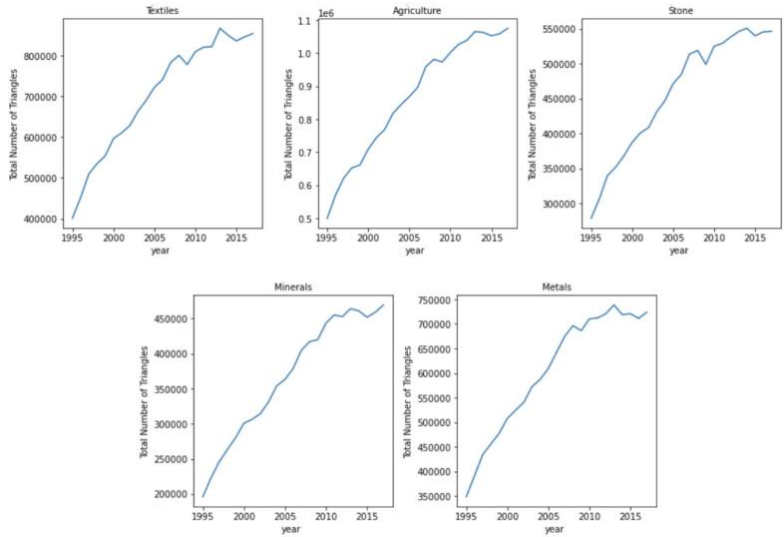

**Fig. 4.** Total number of triangles in the trade networks constructed for the five sectors [Textiles, Agriculture, Stone, Minerals, Metals]. Distributions for the other four sectors were found to be qualitatively similar.

**Table 2.** Pearson's correlations of four metrics against the global GDP series for all ten sectors over the time period 1995-2017.

| Sector | Average Degree | Average Clustering Coefficient | Average Density | Total # Trianges |
|---|---|---|---|---|
| Textiles | 0.918 | 0.607 | 0.917 | 0.945 |
| Agriculture | 0.938 | 0.888 | 0.937 | 0.96 |
| Stone | 0.925 | -0.147 | 0.925 | 0.946 |
| Minerals | 0.94 | 0.595 | 0.941 | 0.962 |
| Metals | 0.914 | -0.208 | 0.914 | 0.936 |
| Chemicals | 0.918 | 0.607 | 0.917 | 0.945 |
| Vehicles | 0.938 | 0.888 | 0.937 | 0.96 |
| Machinery | 0.925 | -0.147 | 0.925 | 0.946 |
| Electronics | 0.94 | 0.595 | 0.941 | 0.962 |
| Other | 0.914 | -0.208 | 0.914 | 0.936 |

Although we do not tabulate the results herein, we also computed the diameter of these international trade networks for each of the nine sectors. Diameters tended to be between 3-4 with the exception of minerals (maximum diameter of 4 was achieved in 2012), and in many cases, a diameter of 2 was already achieved by 1995. This illustrates that, while a trading relation does not exist between every pair of countries in any given sector, an intermediated relation does exist. This is a direct reflection of the degree of globalization that existed in international trade relations in the early part of this century.

## 4   Conclusion

Much has been written in recent years about the benefits and costs of globalization and international trade. It was already well known that global trade and global GDP were highly associated. This paper looked to quantify globalization through a different lens: namely, through a network-theoretic study of trade networks between countries, constructed in a rigorous fashion. Rather than consider a single trade network, we relied on per-sector analyses using data from the Atlas project. We found that there is a heavy correlation between global GDP and multiple network science metrics such as average degree and number of triangles. The average clustering coefficient is less correlated, perhaps because trade networks were already well developed in the mid-1990s when the era of globalization was taking off in the aftermath of the dissolution of the Soviet Union and the liberalization of markets in Asia (most notably, China and India). In general, the average clustering coefficient remains high throughout, but some sectors show higher correlation than others. The data also confirms that globalization and trade relations have been in considerable flux since the 2008 financial crisis.

---

[5] Note that we include two plots in Figure 5 (b), since we found an outlier (East Timorese). The outlier, which occurs around 1999-2000 can be explained by the fact that, following the UN-sponsored act of self determination, East Timorese gained independence from Indonesia in 1999, and consequently became the first new sovereign state of the 21st century.

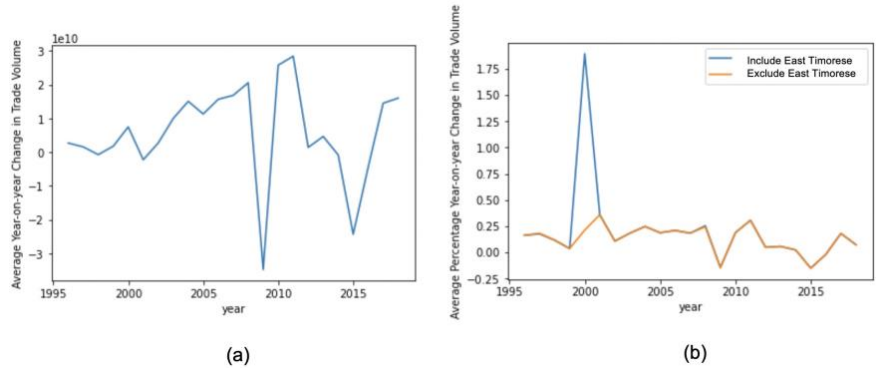

**Fig. 5.** (a) Average change in trade volume (sum of exports and imports), with average taken over countries over the time period 1995-2018. (b) Average change in trade volume (sum of exports and imports), with average taken over countries, compared to the previous year over the time period 1995-2018. Note that, the 1999 East Timorese crisis caused a spike in percentage change in trade volume in 2000.

The data suggests many avenues for future research, not least of which is a study that takes the specific effects of COVID-19 into account. We suspect that there will be significant declines in all measures, and a complete disruption to the trade networks due to the supply chain, and other, bottlenecks caused by the pandemic. Indeed, a closer study of the volume of trade during, and after, this period may help us to quantify both the global and per-country impact of supply chain disruptions through a network science lens.